\documentclass[12pt]{memoir}

\setlrmargins{*}{*}{1}
\checkandfixthelayout

\counterwithout{section}{chapter}

\makeatletter
\renewcommand{\@memmain@floats}{%
  \counterwithin{figure}{section}
  \counterwithin{table}{section}}
\makeatother

\firmlists

\usepackage{mathtools}
\mathtoolsset{mathic} 
\usepackage{amsthm}
\usepackage{amssymb}

\usepackage{enumitem}
\setlist[enumerate]{leftmargin=*}
\setlist[itemize]{leftmargin=*}
\setlist[description]{font=\mdseries\textsf, leftmargin=1.5em}
\usepackage[backref,hyperindex,colorlinks,citecolor=blue]{hyperref}
\usepackage[numbered]{bookmark} 

\usepackage[all]{hypcap} 

\usepackage{float}
\usepackage[algo2e,algosection,tworuled,noend,noline]{algorithm2e}

\newcommand{\customqed}[1]{{\renewcommand{\qedsymbol}{#1}\qed}}
\newcommand{\varqed}{\customqed{\hbox{$\lrcorner$}}}

\theoremstyle{plain}

 \newtheorem{lemma}{Lemma}[section]

 \newtheorem{theorem}{Theorem}

\theoremstyle{definition}

\newtheorem{Definition}[lemma]{Definition}
\newenvironment{definition}{\begin{Definition}}{\varqed\end{Definition}}

 \newtheorem{Notation}[lemma]{Notation}

 \newtheorem{Condition}[lemma]{Condition}

 \newtheorem{Remark}[lemma]{Remark}
 \newtheorem*{Remarknonumber}{Remark} %
 \newtheorem{Remarks}[lemma]{Remarks}

 \newtheorem{Example}[lemma]{Example}
 \newtheorem{Examples}[lemma]{Examples}

 \newcommand{\bP}{\mathbf{P}}
 \newcommand{\bQ}{\mathbf{Q}}
 \newcommand{\bR}{\mathbf{R}}
 \newcommand{\bS}{\mathbf{S}}
 \newcommand{\bT}{\mathbf{T}}

 \newcommand{\bbR}{\mathbb{R}}

 \newcommand{\cA}{\mathcal{A}}
 \newcommand{\cB}{\mathcal{B}}
 \newcommand{\cM}{\mathcal{M}}

\newcommand{\verythinmathskip}{\mskip 1mu}

 \newcommand\Paren[1]{{\left( #1\right)}}
 \newcommand\setOf[2]{\mathopen\{\verythinmathskip#1 : #2\verythinmathskip\mathclose\}}

\newcommand{\df}{\emph}

\newcommand{\capi}{\mathrm{capi}}

\begin{document}

\title{A pair of universal sequence-set betting strategies}
\bookmark[page=1,level=0]{Title}

\author{Tomislav Petrovi\'c}

\maketitle
\thispagestyle{empty}

\begin{abstract}
We introduce the sequence-set betting game, a generalization of An. A. Muchnik's  non-monotonic betting game \cite{Muchnik}. Instead of successively partitioning the infinite binary strings by their value of a bit at a chosen position, as in the non-monotonic game, the player is allowed to partition the strings into any two clopen sets with equal measure. We show that, while there is no single computable sequence-set betting strategy that predicts all non-Martin-L\"of random strings, we can construct two strategies such that every non-Martin-L\"of random string is predicted by at least one of them. 
\end{abstract}
\section{Introduction}
There is a long history of studying the nature of randomness,
even the formal study goes back to at least von Mises \cite{Mises}.
One way of defining randomness is via a kind of a betting game,
a martingale, first used by J. Ville \cite{Ville}. Winning in a game
against infinite string can be viewed as predictability of
a string.
The predictability of a string is formally defined in terms of betting games and computable
strategies. A string is said to be predictable iff there is a computable
strategy that, starting with unit capital, by successive betting,
wins an unbounded amount of capital when betting against the string.
In \cite{Muchnik}, An. A. Muchnik et al., introduce the non-monotonic-betting game. In
this game, the player bets on the bits of the string, each bet consisting
of the index (position) of the bit, wagered amount of capital and
the value of the bit she is betting on. If the value of the bit at the chosen position was guessed correctly the wagered amount is doubled, otherwise it is lost. The strings for which there
is no computable non-monotonic betting strategy that can predict them
are called Kolmogorov-Loveland random, since they both independently proposed non-monotonic inspection of bits in a string \cite{Kolmogorov,Loveland}.
P. Martin-L\"of gives his definition of randomness in terms of computably
enumerable statistical tests in \cite{MartinLof}. His seminal paper
marks a point of departure from the unpredictability paradigm towards
the incompressibility paradigm in defining the random strings.
To define incompressibility, we use the prefix-free variant of Kolmogorov complexity, $K$, see \cite{LiVitanyi}. An incompressible infinite string is such that
for some constant $c$ and all of its prefixes $p$, $K(p)\geq\ell(p)-c$, with $\ell(p)$ denoting the length of the prefix.
It can be shown that Martin-L\"of random strings are precisely the incompressible
ones.
Whether Kolmogorov-Loveland
random sequences are a proper subset of Martin-L\"of random strings
is unknown, and is considered a major open problem in the field of
algorithmic information theory \cite{LiVitanyi,Merkle,Downey,Open,Ambos,Calib}.
A characterization of Martin-L\"of randomness can be given in terms of a certain betting game, we'll call it prefix-betting.
In this game in each iteration the player bets on a segment $s$ of a string extending the known prefix of a string, $p$, and if the string she's playing against starts with the prefix $ps$, she wins some money, if not, the
wagered amount is lost. Since prefix-betting is a fair game, if the
segment she bets on has length $l$, if she is right she wins $2^{l}$
times the wagered amount. It is known that a single computable prefix-betting
strategy can be constructed that predicts all compressible strings \cite{Schnorr}.
Note however, that in the case of a correct guess, we learn something
about the string we are playing against, namely the next segment of
its prefix. But in case we were wrong, we learn very little, only
that it doesn't begin with the prefix we had bet on. This asymmetry
is twofold. Firstly, if the guess was correct, the measure of the
set that we know contains the sequence reduces by a factor of $2^{-l}$,
and if the guess was incorrect it is reduced by a factor of only $(1-2^{-l})$.
Secondly, if we make an infinite number of correct guesses, we learn
all of the bits of the string. On the other hand, if we make an infinite
sequence of wrong guesses, we still might not learn a single bit of the string.
To address the first kind of asymmetry we introduce a generalization of the non-monotonic betting game called the sequence-set betting game.
This is a game where the player, initially starting with the set of
all strings and unit capital, partitions the set of strings into two
clopen sets of equal measure, and bets on one of them.
To see that there is no single computable sequence-set-betting strategy that predicts all compressible strings, at each bet choose the set on which the strategy looses
and obtain a computable sequence of nested sets, each having a measure
of $\frac{1}{2}$ the previous set. The strings in the intersection
of these sets are compressible since we obtained them by a computable
procedure, have measure $0$, and, by construction, the strategy doesn't
predict them. 
On the other hand, we show that it is possible
to construct two such strategies such that every compressible string is
predicted by at least one of them.
\section{Definitions and main theorem}

Our base space is always the space \( \Omega \) of infinite binary sequences,
with the Lebesgue measure \( \lambda \) defined on it.
For a set \( U\subseteq\Omega \) with \( \lambda(U)>0 \) we will define the conditional 
measure \( \lambda_{U} \) on \( U \) as usual by 
\begin{align*}
  \lambda_{U}(A)=\frac{\lambda(A\cap U)}{\lambda(U)}.
 \end{align*}

\begin{definition}[Mass placement]\label{def:mass-placement}
  A \df{mass placement} is a triple \( P=(U,\cA,\mu) \) where
\( \cA \) is a finite partition of the clopen \df{ground set}
\( U\subseteq\Omega \) into clopen sets,
with a mass function \( \mu:\cA\to\bbR_{+} \) defined on them.
We will call \( \cA \) the \df{partition} of \( P \).
The above mass placement is called \df{atomic} if its partition 
consists of a single element.

The mass function is extended to all unions of elements of  \( \cA \): 
\( \mu(A_{1}\cup\dots\cup A_{i}) = \mu(A_{1})+\dots+\mu(A_{i}) \).
The \df{minimum capital} of \( P \) is
\begin{align*}
  \capi(P)=\min_{A\in\cA}\frac{\mu(A)}{\lambda(A)}.
 \end{align*}
The \df{grain bound} of \( P \) is \( \max_{A\in\cA}\lambda(A) \).

We will say that a mass placement \( Q=(U,\cB,\nu) \) is a \df{refinement} of
a mass placement \( P=(U,\cA,\mu) \), writing \( Q\preceq P \),
if each \( B\in \cB \) is subset of some \( A\in \cA \), further
\( \mu(A) = \nu(A) \) for all \( A\in\cA \).

\end{definition}

\begin{definition}[Grid]\label{def:grid}
Consider a pair of mass placements \( \bP=(P^{0},P^{1}) \),
\( P^{j}=(U,\cA^{j},\mu^{j}) \), \( j=0,1 \) 
with the same ground set \( U=U_{\bP} \).
We will call such an object a \df{grid}.
In general, for any grid \( \bP \) we will denote its ground set by \( U_{\bP} \),
its partitions by \( \cA^{j}_{\bP} \),
and its mass functions by \( \mu^{j}_{\bP} \) for \( j=0,1 \).
So
 \begin{align*}
   \bP = (P^{0},P^{1}),\ P^{j} = (U_{\bP},\cA^{j}_{\bP},\mu^{j}_{\bP}),\ j=0,1.
 \end{align*}
Let \( \capi(\bP)=\min(\capi(P^{0}),\capi(P^{1})) \).
The \df{grain bound} of \( \bP \) is the maximum of the grain bounds of \( P^{0} \) and \( P^{1} \).
A grid is \df{atomic} if both of its elements are.
We will say that grid \( \bP \) is  a \df{refinement} of grid \( \bQ \)
and write \( \bP\preceq \bQ \) if \( P^{j}\preceq Q^{j} \) for \( j=0,1 \).

A set \( S \) is \df{compatible} with the grid \( \bP \)
if it is the union of sets of the form \( E^{0}\cap E^{1} \) with \( E^{j}\in\cA^{j}_{\bP} \).
\end{definition}

\begin{definition}[Sum]\label{def:sum}
Let \( \bP,\bQ \) be grids with disjoint ground sets.
Consider the mass placement \( \bR \) with
\begin{align*}
             U_{\bR} &= U_{\bP}\cup U_{\bQ},
\\   \cA^{j}_{\bR} &= \cA^{j}_{\bP}\cup\cA^{j}_{\bQ},
 \end{align*}
with \( \mu^{j}_{\bR}(E) =\mu^{j}_{\bP}(E) \) for \( E\in\cA^{j}_{\bP} \) 
and \( \mu^{j}_{\bQ}(E) \) for \( E\in\cA^{j}_{\bQ} \), \( j=0,1 \).
We will denote \( \bR=\bP+\bQ \).
Of course, \( \capi(\bR) =\min(\capi(\bP),\capi(\bQ)) \).
A sum of three or more elements is defined similarly.
A grid \( \bP \) is called \df{diagonal} if \( \cA^{0}_{\bP}=\cA^{1}_{\bP} \),
or equivalently, if it is a sum of atomic grids.
\end{definition}

\begin{definition}[Orthogonality]\label{def:orthogonal}
Somewhat dual to diagonality is the concept of orthogonality.
The grid \( \bP \) is called
\df{orthogonal} if for all \( A\in\cA^{0}_{\bP} \) 
and \( B\in\cA^{1}_{\bP} \) we have
\begin{align*}
   \lambda_{U}(A\cap B) = \lambda_{U}(A)\lambda_{U}(B).
 \end{align*}
(This means that the two algebras, defined by \( \cA^{0} \) and \( \cA^{1} \) are
conditionally independent on \( U_{\bP} \).)
  Let \( \bP \) be a grid.
Let \( j\in\{0,1\} \).
We can replace an arbitrary element \( A\in\cA^{j}_{\bP} \) 
with \( 2 \) disjoint subsets \( A=A_{1}\cup A_{2} \) with
\( \lambda(A_{i}\cap B)=\lambda(A\cap B)/2 \) for all \( B\in\cA^{1-j}_{\bP} \),
and choose \( \mu^{j}_{\bP}(A_{i})=\mu^{j}_{\bP}(A)/2 \).
The new grid will be said to be obtained by an \df{even split} from \( \bP \).
Of course it has the same ground set and min and max capital and  as the original one.
Also, splitting conserves orthogonality.
\end{definition}
\begin{sloppypar}
\begin{definition}[Mass placement test]\label{def:test}
We call a mass placement 
\({(\Omega,\{\Omega\},\mu(\Omega)=1)}\) a \df{unit mass placement} 
and an infinite sequence of mass placement refinements 
\(T=T_0\succeq T_1\succeq...\) a \df{mass placement test} iff \(T_0\) is a unit mass placement.

Denote with \(\cA_T=\bigcup_{i}\cA_{T_i}\). 
The mass function \(\mu_T:\cA_T\to\bbR_+\) is defined by the mass placements \(T_i\), 
for \(A\in \cA_{T_i}\), \(\mu_T(A)=\mu_{T_i}(A)\). 
The mass function is extended to all unions of disjoint \(\cB\subseteq\cA_T\): 
\(\mu_T(\bigcup\cB)=\underset{B\in\cB}{\sum}\mu_T(B)\). 
For brevity, we drop \(\bigcup\) and write just \(\mu_T(\cB)\).

The \df{capital} of \(A\in\cA_T\) is
\begin{align*}
c(A)=\frac{\mu_T(A)}{\lambda(A)}
\end{align*}
For \(\alpha\in\Omega\), denote with \(T_\alpha\), a set of elements of \(\cA_T\) 
that contain \(\alpha\).  
We say that \(\alpha\) \df{fails} the test \(T\) iff
\begin{align*}
\underset{A\in T_\alpha}{\sup}c(A)=\infty
\end{align*}

We call a mass placement test a \df{granular test} iff the grain bound of \(T_i\) goes to zero in the limit.
\end{definition}
  \end{sloppypar}

\begin{definition}[Sequence-set betting strategy]\label{def:ssb}
A mass placement test \(S\) is a \df{sequence-set betting strategy} if for all \(A\in\cA_{S_i}\), \(\lambda(A)=2^{-i}\). 
\end{definition}

Clearly, a sequence set betting strategy is a granular test.
\begin{definition}[Martin-L\"of test]
A nested sequence of open sets \(N= N_1 \supseteq N_2 \supseteq\dots\)  
with \(\lambda(N_i)\leq 2^{-i}\) and \(N_i\) computably enumerable 
uniformly in \( i \) 
is called a \df{Martin-L\"of test}. 
An infinite binary sequence \(\alpha\) fails the test \(N\) iff \(\alpha\in\bigcap_{i\in\mathbb{N}}N_i\). A Martin-L\"of test \(M\) is called universal if every \(\alpha\) that fails some Martin-L\"of test \(N\) also fails \(M\).
\end{definition}
It is well-known that there is an universal Martin-L\"of test \cite{MartinLof,LiVitanyi}.
\begin{theorem}\label{mainThm}
For a 
Martin-L\"of test \(M\), there are two computable sequence-set betting strategies
\(S^0,S^1\) such that for every infinite binary sequence \(\alpha\) that fails \(M\), \(\alpha\) 
fails at least one of \(S^0,S^1\).
\end{theorem}

\section{Proof}

\begin{sloppypar}
\begin{lemma}\label{lem:key-lemma}
Consider a diagonal grid \( \bP \) with \( c=\capi(\bP) \),
and a sequence of disjoint clopen sets \( E_{1},E_{2},\dots \), 
 \( E_{i}\subseteq U_{\bP} \), \( \sum_{i} \lambda(E_{i})\le p^{2} \lambda(U_{\bP}) \) for 
a certain \( p<1 \).
Let \( K>1 \) be a constant with \( Kp<1 \).
We can compute a sequence of disjoint diagonal grids \( \bQ_{1},\bQ_{2},\dots \) with
\( U_{\bQ_{i}}=E_{i} \), and
for each \( n>0 \) a mass placement \( \bR_{n} \) disjoint from 

\( \bQ_{1},\bQ_{2},\dots,\bQ_{n} \), 
an even splitting
\( \bP_{n} \) of \( \bP \) and even splittings \( \bQ_{i,n} \)
of \( \bQ_{i} \) such that the following holds:
\begin{align*}
 \bQ_{1,n}+\dots+\bQ_{n,n}+\bR_{n} &\preceq \bP_{n},
\\ \bQ_{n+1}+\bR_{n+1} &\preceq \bR_{n},
 \end{align*}
and for each \( i \), for each atomic part \( \bT \) of \( \bQ_{i} \) we have
\begin{align}\label{eq:K-max} 
                            \max(\capi(T^{0}),\capi(T^{1}))&= c K,
\\\label{eq:K-min}    \min(\capi(T^{0}),\capi(T^{1})) &=\capi(\bT)\ge c',
 \end{align}
where \( c'=c(1-pK) \).
Also the grain bound of \( \bP_{n} \) is \( \le 2^{-n} \).
\end{lemma}
The \( 2^{-n} \) grain bound is arbitrary, it only matters that it converges to 0 constructively.
 \end{sloppypar}
\begin{proof}
We construct the grids \( \bQ_{i} \) and \( \bR_{n} \) explicitly and recursively.
Suppose that \( \bQ_{i} \), \( \bR_{i} \) and \( \bP_{i} \)
have been defined already for all \( i<n \).
Let \( \bR=\bR_{n-1} \), \( \bS = \bR_{n} \) (this is still to be defined).
Denote 
\begin{align*}
V_{n} = E_{1}\cup\dots\cup E_{n}.
 \end{align*}
Of course, \( U_{\bS}=U_{\bP}\setminus V_{n} \).
It is possible to replace \( \bP_{n-1} \) with a \( \bP_{n} \) obtained by repeated
even splittings in such a way that it becomes compatible with the clopen set \( E_{n} \)
(and also satisfies the grain bound).
We  obtain each \( \bQ_{i,n} \) from \( \bQ_{i} \) and \( \tilde\bR \) from \( \bR \) similarly.
Note that since \( \bP_{n-1} \) was orthogonal, so is \( \bP_{n} \).
With these splittings we  can achieve the following: for each \( j=0,1 \), for
each \( B\in \cA^{j}_{\bP_{n}} \), and \( i<n \) we have
\begin{align*}
   B\cap E_{i}\in\cA^{j}_{\bQ_{i,n}},\ 
   B\setminus V_{n-1}\in\cA^{j}_{\tilde\bR}. 
 \end{align*}
Now, as \( E_{n} \) is compatible with \( \bP_{n} \), 
it can be written as a disjoint union \( E_{n}=F_{1}\cup\dots\cup F_{m} \), where
each \( F_{k} \) has the form \( B^{0}\cap B^{1} \), with \( B^{j}\in\cA^{j}_{\bP_{n}} \).
The (diagonal) grid \( \bQ_{n} \) will be a sum of atomic grids \( \bT_{1}+\dots+\bT_{m} \),
with \( U_{\bT_{k}}=F_{k} \).
Of course, \( U_{\bS}=U_{\bR}\setminus E_{n} \), and 
\begin{align*}
\cA^{j}_{\bS} = \setOf{A\setminus V_{n}}{A\in \cA^{j}_{\bP_{n}}}
 = \setOf{B\setminus E_{n}}{B\in\cA^{j}_{\tilde\bR}}.
 \end{align*}
The interesting part is the definition of the new mass functions
\( \mu^{j}_{\bT_{k}} \) and \( \mu^{j}_{\bS} \).
Note that there is a common refinement \( \hat\bP_{n} \) of
\( \bP_{n} \) and \( \bQ_{1,n},\dots,\bQ_{n,n} \), \( \bR_{n} \).
(We use \( \hat\bP_{n} \) only for the calculations in this proof, not the actual
construction.)
Wherever \( \mu^{j}_{\bT_{k}} \) and \( \mu^{j}_{\bS} \) 
are defined, they are identical to \( \mu^{j}_{\hat\bP_{n}} \).
Therefore from now on, we omit the subscript of \( \mu^{j} \) when it signifies
\( \mu^{j}_{\hat\bP_{n}} \) (which is essentially in all cases).
For \( j=0,1 \), for each \( B\in \cA^{j}_{\tilde\bR} \), 
we need to redistribute the mass \( \mu^{j}(B) \)
into \( B\cap V_{n} \) and \( B\setminus V_{n} \).
The mass  assigned to \( B\cap V_{n} \) will be distributed among all the \( F_{k} \)
with \( F_{k}\subseteq  B \), and the mass remaining is given to \( B\setminus V_{n} \).
Let
\begin{align*}
\cM=\setOf{A\in\cA^{0}_{\bP_{n}}}{\lambda(A\cap V_{n})> p\lambda(A)}.
 \end{align*}
First we determine the values of \( \mu^{0}(\cdot) \).
Suppose \( A\notin\cM \).
Recalling \( A\cap U_{\bR} =  (A\setminus V_{n})\cup (A\cap E_{n}) \), a disjoint union, 
for \( F_{k}\subseteq A \) we set 
\begin{equation}\label{eq:mu0}
\begin{aligned}
    \mu^{0}(F_{k}) &=
  \begin{cases}
       c K \lambda(F_{k}) & \text{if } A\notin \cM,
   \\ c'\lambda(F_{k}) & \text{otherwise,}
  \end{cases}
\\ \mu^{0}(A\setminus V_{n}) &= \mu^{0}(A\cap U_{\bR})-\mu^{0}(A\cap E_{n}).
 \end{aligned}
  \end{equation}
In other words, for each \( F_{k}\subseteq A \), if \( A\notin \cM \) then
we satisfy the requirement~\eqref{eq:K-max} by
by \( \capi(T^{0}_{k})=c K \), otherwise we satisfy~\eqref{eq:K-min}.

Now we determine the values of \( \mu^{1}(\cdot) \).
Let \( B\in\cA^{1}_{\bP_{n}} \).
Let \( F_{k}=A\cap B \) with \( A\in\cA^{0}_{\bP_{n}} \).
We set
\begin{equation}\label{eq:mu1}
\begin{aligned}
    \mu^{1}(F_{k}) &=
                               \begin{cases}
  c K \lambda(F_{k}) &\text{if }  A\in \cM,
\\ c'\lambda(F_{k})                 &\text{otherwise,}
                               \end{cases}
\\ \mu^{1}(B\setminus V_{n}) &= \mu^{1}(B\cap U_{\bR})-\mu^{1}(B\cap E_{n}).
 \end{aligned}
\end{equation}
These definitions satisfy conditions~\eqref{eq:K-min}, \eqref{eq:K-max} by design;
what remains to prove is that they provide
\( \mu^{j}(A\setminus V_{n})\ge 0 \)  for \( A\in\cA^{j}_{\bP_{n}} \).
We will actually prove the stronger inequality
\( \mu^{j}(A\setminus V_{n})\ge c'\lambda(A\setminus V_{n}) \).

Consider the case \( j=0 \), and first the case \( A\notin\cM \).
We will prove, inductively, the stronger statement
\begin{equation}\label{eq:K-min-3}
  \mu^{0}(A\setminus V_{n}) 
    \ge c \Paren{1-K \frac{\lambda(A\cap V_{n})}{\lambda(A)}}\lambda(A)
    = c \lambda(A)- cK\lambda(A\cap V_{n}).
\end{equation}
If \( A\notin\cM=\cM_{n} \) then for all \( i<n \), for 
\( A'\supseteq A \) with \( A'\in \cA^{0}_{\bP_{i}} \) we had \( A'\notin\cM_{i} \).
Therefore by inductive assumption, for \(i=n-1\),
\begin{align*}
  \mu^{0}(A'\setminus V_{n-1})\ge  c\lambda(A')- cK\lambda(A'\cap V_{n-1}).
 \end{align*}
This linear inequality is conserved by the splittings, therefore also 
\begin{align}\label{eq:non-M-ind}
  \mu^{0}(A\setminus V_{n-1})\ge c\lambda(A)-c K\lambda(A\cap V_{n-1}).
 \end{align}
Now by definition~\eqref{eq:mu0}
\begin{align*}
 \mu^{0}(A\setminus V_{n}) &=\mu^{0}(A\setminus V_{n-1}) -\mu^{0}(A\cap E_{n})
= \mu^{0}(A\setminus V_{n-1}) - c K \lambda(A\cap E_{n}),
 \end{align*}
which together with~\eqref{eq:non-M-ind} proves~\eqref{eq:K-min-3}.

Now consider the case \( A\in\cM \).
Let \( l \) be the last \( i<n \) where for
\( A'\supseteq A \) with \( A'\in \cA^{0}_{\bP_{i}} \) we had \( A'\notin\cM_{i} \).
If there is no such \( i \) then set \( l=0 \), this case is trivial.
Then by the above proof, for such \( A' \) we have
\begin{align*}
  \mu^{0}(A'\setminus V_{l-1})\ge c\lambda(A')-c K\lambda(A'\cap V_{l-1}) 
  \ge c'\lambda(A'\cap V_{l-1}).
 \end{align*}
From now on the process for all \( i>l \) will either split evenly or assign
\( \mu^{0}(A\cap E_{i})=c'\lambda(A\cap E_{i}) \) and 
\( \mu^{0}(A\setminus V_{i}) = 
\mu^{0}(A\setminus V_{i-1})-c'\lambda(A\cap E_{i}) \).
Both kinds of step conserve the lower bound \( c'>0 \) on capital.

Now consider the case \( j=1 \);
we want to prove \( \mu^{1}(B\setminus V_{n})\ge 0 \)
for \( B\in\cA^{1}_{\bP_{n}} \).
Let \( M=\bigcup\cM \).
We claim
\begin{align}\label{eq:B-M}
\lambda(B\cap M)\le p\lambda(B).
 \end{align}
For this, we will use the orthogonality of \( \bP_{n} \) and the assumption
\( \lambda(V_{n})\le p^{2}\lambda(U_{\bP}) \) of the lemma.
For every \( A\in\cM \) we have by definition of \( \cM \),
\( \lambda(A\cap V_{n})>p\lambda(A) \).
Let \( u=\lambda(U_{\bP_{n}}) \). 
Then
\begin{align*}
  p^{2}u\ge \lambda(V_{n}) \ge \lambda(M\cap V_{n}) > p\lambda(M),
 \end{align*}
implying \( \lambda(M)<p u \).
By orthogonality,
\begin{align*}
\lambda(B\cap M)/u = (\lambda(B)/u)(\lambda(M)/u) \le p \lambda(B)/u,
 \end{align*}
implying~\eqref{eq:B-M}.
By definition~\eqref{eq:mu1}, and using~\eqref{eq:B-M},
\begin{align*}
\mu^{1}(B\cap M\cap V_{n})\le c K \lambda(B\cap M\cap  V_{n})\le c K p\lambda(B),
 \end{align*}
hence
\begin{align*}
\mu^{1}(B\setminus (M\cap V_{n})) &\ge \mu^{1}(B)-c K \lambda(B\cap M)
\\ & \ge c\lambda(B)(1-Kp) = c'\lambda(B) \ge c'\lambda(B\setminus (M\cap V_{n})).
 \end{align*}
By definition~\eqref{eq:mu1},
\( \mu^{1}(B\cap V_{n}\setminus M)=c'\lambda(B\cap V_{n}\setminus M) \),
hence 
\begin{align*}
\mu^{1}(B\setminus V_{n})\ge
c'\lambda(B\setminus (M\cap V_{n})) - c'\lambda(B\cap V_{n}\setminus M)
= c'\lambda(B\setminus V_{n}),
 \end{align*}
finishing the  proof.
 \end{proof}

\begin{sloppypar}
\begin{lemma}\label{lem:granular}
For any Martin-L\"of test \(N\) we can computably construct two granular tests \(P^0,P^1\) such that every infinite 
binary sequence \(\alpha\) that fails \(N\), \(\alpha\) fails \(P^0\) or \(\alpha\) fails \(P^1\).
\end{lemma}
 \end{sloppypar}
 \begin{proof}
We construct the granular tests explicitly and recursively.

For some \( i \), suppose 
that \(\bQ\) is a diagonal grid with \(U_\bQ=H\) for some \(H\in N_i\). 
Recall that a diagonal grid is a sum of atomic grids and let
\(\bT_1,\dots,\bT_n\) be atomic grids such that
\(\bT_1+\dots+\bT_n=\bQ\). Let \(\bT_k=(T^0_k,T^1_k)\)
Denote
\begin{align*}
m(\bQ)=\min_k\max(\capi(T^0_k),\capi(T^1_k))
\end{align*} 
 and 
let \(c=\capi(\bQ)\), \(K=2m(\bQ)/c\) and \( r \) such that \(2^{-r}\leq 1/(cK)^2\).
Let \(E_1,E_2,\dots\) be an enumeration of elements of \(N_{r}\cap H\).
By Lemma~\ref{lem:key-lemma}, 
there is a sequence of disjoint diagonal grids \(Q_{T,\bQ}= \bQ_{1},\bQ_{2},\dots \) 
and a sequence of mass placements \(R_{T,\bQ}=\bR_1,\bR_2,\dots\), such that
\(\bR_i\) is disjoint from \(\bQ_1,\dots,\bQ_i\), 
\(\bQ\succeq\bQ_1+\dots+\bQ_i+\bR_i\),
\( U_{\bQ_{j}}=E_{j} \), and
\begin{align*}
m(\bQ_j)= K c = 2m(\bQ).
\end{align*}

Denote by \(\bP_0\) a grid consisting of a pair of unit mass placements.
Let \(\bQ=\bP_0\), denote with \(\bQ_{i_1,\dots,i_n,k}\) 
the \(k\)-th element of \(Q_{T,\bQ_{i_1,\dots,i_n}}\)
and with \(\bR_{i_1,\dots,i_n,k}\)
the \(k\)-th element of \(R_{T,\bQ_{i_1,\dots,i_n}}\). Denote by
\begin{align*}
\bP_k=\underset{i_1+\dots+i_n=k}{\sum}\bQ_{i_1,\dots,i_n}+\bR_{i_1,\dots,i_n}
\end{align*}
a sum of disjoint grids. We have that \(\bP_k\succeq\bP_{k+1}\) since
 \(\bR_{i_1,\dots,i_n}\succeq\bQ_{i_1,\dots,i_n+1}+\bR_{i_1,\dots,i_n+1}\)
 and \(\bQ_{i_1,\dots,i_n}\succeq\bQ_{i_1,\dots,i_n,1}+\bR_{i_1,\dots,i_n,1}\).
We have obtained a sequence of grid refinements
\(\bP_0\succeq \bP_1\succeq\dots \).
Note that since \(T\) is a test, \(N_0=\{\Omega\}\) and \(U_{P_0}=\bigcup N_0\).
Then for every sequence of clopen sets \(E=E_0\supseteq E_1\supseteq\dots\), \(E_i\in N_i\),
 there is an infinite subsequence \(E'=E'_0\supseteq E'_1\supseteq\dots\),
 chosen in the following way:
 let \(E'_0=E_0\) and if for some \(i_1,\dots,i_n\), \(E'_i=U_{\bQ_{i_1,\dots,i_n}}\),
 then there is some \(k\) 
 such that \(U_{\bQ_{i_1,\dots,i_n,k}}\in E\). 
 Let \(E'_{i+1}=U_{\bQ_{i_1,\dots,i_n,k}}\).
Then for every \(E'_i\) there is a 
refinement \(\bP_{i'}\) that contains a diagonal grid \(\bQ\) with \(U_\bQ=E'_i\) and \(m(\bQ)=2^i\). 
Then for \(\alpha\in\bigcap_{i\in\mathbb{N}}E_i\) for at least one \(j\in\{0,1\}\) we'll 
have \(\underset{A\in\cA^j_\bP,\alpha\in A}{\sup}\mu^j(A)/\lambda(A)=\infty\).  
Let \(P^0=P_0^0\succeq P_1^0\succeq \dots\) and \(P^1=P_0^1\succeq P_1^1\succeq \dots\) . 
We have that \(P^0,P^1\) satisfy the lemma.
\end{proof}

\begin{sloppypar}
\begin{lemma}\label{lem:ssb}
For any granular test \(P\) we can computably construct a sequence-set betting strategy \(S\)
such that for every infinite binary sequence \(\alpha\) that fails \(P\), \(\alpha\) fails \(S\).
\end{lemma}
 \end{sloppypar}
\begin{proof}
In this proof \( \subset \) means strict inclusion.
Let  \( h>1 \),  \( A\in \cA_{S} \) for some sequence-set betting strategy \(S\) and \(\cB\) a finite, disjoint subset of \(\cA_P\). 
We'll say \(A\) \df{corresponds} to \(\cB\), and write \(A\sim\cB\) iff \(A\subseteq\bigcup\cB\) 
and two conditions are satisfied:
\begin{equation}\label{eq:eqmass}
\mu_P(\cB)<h\mu_S(A)
\end{equation}
\begin{equation}\label{eq:capped}
\forall B\in\cB~\underset{B\subset B'\in\cA_P}{\max}c(B')<
2h\underset{A\subseteq A'\in\cA_{S}}{\max}c(A').
\end{equation}

We'll explicitly and recursively construct \(S\) with the property
\begin{equation}\label{eq:scorrb}
\forall A\in\cA_S~\exists\cB\subseteq\cA_P~ A\sim\cB
\end{equation}
We claim that any sequence-set betting strategy with the property ~\eqref{eq:scorrb} satisfies the theorem. 
We prove the claim by contradiction. 
Suppose there is some infinite binary sequence \(\alpha\) that fails \(P\) and passes \(S\). 
For any \(\alpha\) the following statement is either true or false:
\begin{equation}\label{eq:infcorr}
 \forall B\in P_\alpha~\forall^\infty A\in S_\alpha~\forall\cB\subseteq\cA_P~ A\sim\cB\Rightarrow B\not\in\cB
\end{equation}
Suppose~\eqref{eq:infcorr} is false. 
Then there is some \(B\in P_\alpha\) such that for infinitely many \(A\in S_\alpha\) there is some \(\cB\subseteq\cA_P\) that contains \(B\) and \(A\sim\cB\). We cannot have that \(\mu_P(B)=0\) as this is contrary to the assumption that \(\alpha\) fails \(P\). If \(\mu_P(B)>0\) then since \(\underset{A\in S_\alpha}{\lim}\lambda(A)=0\), by~\eqref{eq:eqmass} we have that \(\underset{A\in S_\alpha}{\sup}c(A)=\infty\) contrary to the assumption that \(\alpha\) passes \(S\). 

Suppose~\eqref{eq:infcorr} is true. 
Note that for any two elements of \( \cA_{P} \) they are either disjoint or one contains the other.
Then for every \(B\in P_\alpha\) and \(A\in S_\alpha\) let \( A\sim \cB_{A} \) for
some \(\cB_{A}\subseteq\cA_P\) .
For almost all \( A \), then \( B\not\in \cB_{A} \).
But \( \alpha\in B' \) for some \( B'\in\cB_A \), and then either \( B\subset B' \) or \( B'\subset B \).
Since there are finitely many \(B'\supset B \) in \(\cA_P\), and they can be only in \( \cB_{A} \) for finitely many \(A\)'s,
for all but finitely many \( A \) we have \( B'\subset B\). 
From~\eqref{eq:capped} there is some \(A'\supseteq A \) with \( c(B)<2 h c(A')\)  and 
by the assumption that \(\alpha\) fails \(P\) 
we have that \(\underset{A\in S_\alpha}{\sup}c(A)=\infty\), contrary to the 
assumption that \(\alpha\) passes \(S\).
This proves the claim.

Now to the construction of \( S \).
Since both \(P\) and \(S\) are tests, we have that \(S_0\) corresponds to \(\{P_0\}\).
Let \( \cA \) be the part of \( \cA_{S} \) already constructed, and assume it satisfies~\eqref{eq:scorrb}.
Then for all \(A\in\cA \) there is 
some \(\cB\subseteq\cA_P\) with \(A\sim\cB\). 
Denote 
\(C_\cB=\{B\in\cB : c(B)<2h\underset{A\subseteq A'\in\cA_S}{\max}c(A')\}\). 
Then \(  C_{\cB}\ne\emptyset \), since
\begin{align}\label{eq:CB-lb}
 \lambda(\cB\setminus C_{\cB})\le \frac{1}{2}\lambda(A)
 \end{align}
is implied by~\eqref{eq:eqmass}.

Let \( k \) be such that \(B\in C_\cB\), \(B\in\cA_{P_{k-1}}\) and \(B\not\in\cA_{P_k}\). 
Replacing \(B\) with it's partitions in \(\cA_{P_k}\) we obtain \(\cB'\).
We have \(A \sim \cB'\) since \(\bigcup\cB=\bigcup\cB'\), satisfying~\eqref{eq:eqmass}, 
and~\eqref{eq:capped} is satisfied for elements \( B'\ne B \) of \( \cB' \)
due to~\eqref{eq:capped} for \( \cB \), and for \( B \) itself is due to \(B\in C_\cB\). 
Applying iteratively, since \(P\) is granular, for any \(\epsilon\) we can obtain \(\cB'\) such that for any \(B\in C_{\cB'}\), \(\lambda(B)<\epsilon\). 
Note that there is at least one \(M\in C_{\cB'}\) with \(c(M)\leq c(\cB')\) 
and \( A\cap M\ne \emptyset \)
since for all \(B\in\cB'\setminus C_{\cB'}\), \(c(B)\geq c(\cB')\). 
Note that~\eqref{eq:CB-lb} holds also for \( \cB' \).
We now partition \(\cB'\) into three sets \(\cB'_0,\{M\},\cB'_1\), having 
\(\cB'\setminus C_{\cB'}\subseteq\cB'_0\), and distributing the elements of \(C_{\cB'}\setminus\{M\}\) 
among \(\cB'_0,\cB'_1\) so that for \(j\in\{0,1\}\), 
\(\lambda(\bigcup\cB'_j\cap A)\leq \frac{1}{2}\lambda(A)\). 
Furthermore, we can partition \(M\) into \(M_0,M_1\) so that the sets \(A_0,A_1\)
\begin{equation*}
 A_j=A\cap(\bigcup\cB'_j\cup M_j) \text{ for } j=0,1
\end{equation*}
 have \(\lambda(A_j)=\frac{1}{2}\lambda(A)\). Let \(\cB_j=\cB'_j\cup\{M\}\) and
 \begin{equation*}
 \mu_S(A_j)=\frac{\mu_P(\cB_j)}{\mu_P(\cB_0)+\mu_P(\cB_1)}\mu_S(A)
 \end{equation*}  We have \(\mu_P(\cB_0)+\mu_P(\cB_1)=\mu_P(\cB')+\mu_P(M)<(1+\epsilon)\mu_P(\cB')\). Let \(h'=\mu_P(\cB')/\mu_S(A)\). Then \(\mu_P(\cB_j)<h'(1+\epsilon)\mu_S(A_j)\), from~\eqref{eq:eqmass} \(h'<h\), and by choosing small enough \(\epsilon\),~\eqref{eq:eqmass} is satisfied for \(A_j,\cB_j\). We have that~\eqref{eq:capped} is satisfied for \(A_j,\cB_j\) since it is satisfied for \(A,\cB'\). We have shown from the inductive assumption that \(A\sim\cB\) for some \(\cB\subseteq\cA_P\), we obtain two subsets of \(\cA_P\), \(\cB_0,\cB_1\), a partition of \(A\) into two sets of equal measure, \(A_0,A_1\), and their mass assignments  such that \(A_0\sim\cB_0\) and \(A_1\sim\cB_1\), proving that \(S\) has property~\eqref{eq:scorrb}.

\end{proof}

\begin{proof}[Proof of Theorem \ref{mainThm}]
By Lemmas~\ref{lem:granular} and~\ref{lem:ssb} for any Martin-L\"of test, and in particular the universal one, \(M\), we can computably construct two sequence-set betting strategies \(S^0,S^1\) such that every infinite binary sequence that fails \(M\) also fails at least one of \(S^0,S^1\).
\end{proof}

\section*{Acknowledgements}
I thank Wolfgang Merkle for our discussions and support, Jason Rute for his suggestions on improving the readability, Mark Lemay for proof reading and corrections, and Peter G\'acs for his most generous help in improving and streamlining the proof and helping form the final version of the paper.


\begin{thebibliography}{}
\bibitem{LiVitanyi}Ming Li, Paul M. B. Vit\'anyi: An Introduction to
Kolmogorov Complexity and Its Applications, Third Edition. Texts in
Computer Science, Springer 2008, ISBN 978-0-387-33998-6, pp. i-xxiii,
1-790

\bibitem{Muchnik}Andrei A. Muchnik, Alexei L. Semenov, Vladimir A.
Uspensky: Mathematical Metaphysics of Randomness. Theor. Comput. Sci.
207(2): 263-317 (1998)

\bibitem{Schnorr}C.P. Schnorr. Zuf\"alligkeit und Wahrscheinlichkeit; Eine algorithmische Begr\"undung der Wahrscheinlichkeitstheorie, volume 218 of Lect. Notes Math. Springer-Verlag, Berlin, 1971.

\bibitem{Mises}R. von Mises. Grundlagen der Wahrscheinlichkeitsrechnung.
Mathematische Zeitschrift, 5:52\textendash 99, 1919. {[}xviii, 229{]}



\bibitem{Ville}J. Ville. Etude Critique de la Notion de Collectif.
Gauthier-Villars, Paris, 1939.

\bibitem{Kolmogorov}A.N. Kolmogorov. On tables of random numbers.
Sankhya , The Indian Journal of Statistics, Ser. A, 25:369\textendash 376,
1963.

\bibitem{Loveland}D.W. Loveland. A new interpretation of von Mises\textquoteright{}
concept of a random sequence. Z. Math. Logik und Grundlagen Math.,
12:279\textendash 294, 1966.

\bibitem{MartinLof}P. Martin-L\"of. The definition of random sequences.
Inform. Contr., 9:602\textendash 619, 1966.


\bibitem{Merkle}Wolfgang Merkle, Joseph S. Miller, Andr\'e Nies, Jan
Reimann, Frank Stephan: Kolmogorov-Loveland randomness and stochasticity.
Ann. Pure Appl. Logic 138(1-3): 183-210 (2006)

\bibitem{Downey}Rodney G. Downey, Denis R. Hirschfeldt: Algorithmic
Randomness and Complexity. Theory and Applications of Computability,
Springer 2010, ISBN 978-0-387-95567-4, pp. 1-766

\bibitem{Open}Joseph S. Miller, Andr\'e Nies: Randomness and Computability:
Open Questions. Bulletin of Symbolic Logic 12(3): 390-410 (2006)

\bibitem{Ambos}K. Ambos-Spies and A. Ku\v{c}era. Randomness in computability
theory. In P. Cholak, S. Lempp, M. Lerman, and R.A. Shore, editors,
Computability Theory and Its Applications: Current Trends and Open
Problems, vol- ume 257 of Contemporary Mathematics, pages 1\textendash 14.
American Math. Society, 2000.

\bibitem{Calib}Rodney G. Downey, Denis R. Hirschfeldt, Andr\'e Nies,
Sebastiaan Terwijn: Calibrating Randomness. Bulletin of Symbolic Logic
12(3): 411-491 (2006)



\end{thebibliography}
\end{document}